\def\ref{\par\noindent\hang}

\def\spose#1{\hbox to 0pt{#1\hss}}
\def\approxlt{\mathrel{\spose{\lower 3pt\hbox{$\sim$}}
        \raise 2.0pt\hbox{$<$}}}
\def\approxgt{\mathrel{\spose{\lower 3pt\hbox{$\sim$}}
        \raise 2.0pt\hbox{$>$}}}

\def\multleft#1{\hbox to size{\vbox {\halign {\lft{##}\cr #1}}\hfill}\par}
\def\multright#1{\hbox to size{\vbox {\halign {\rt{##}\cr #1}}\hfill}\par}

\def\degmark{^\circ}
\def\s{\hbox{\phantom{5}}}

\def\boxit#1{\vbox{\hrule\hbox{\vrule\kern3pt\vbox{\kern3pt
          #1 \kern3pt}\kern3pt\vrule}\hrule}}

\def\cm{{\rm\thinspace cm}}

\def\erg{{\rm\thinspace erg}}
\def\eV{{\rm\thinspace eV}}

\def\keV{{\rm\thinspace keV}}

\def\s{{\rm\thinspace s}}

\def\ergpcmsqps{\hbox{$\erg\cm^{-2}\s^{-1}\,$}}

\def\ergps{\hbox{$\erg\s^{-1}\,$}}

\def\pcmsq{\hbox{$\cm^{-2}\,$}}

\documentclass{article}

\usepackage{psfig}
\usepackage{aaspp4}

\makeatletter
\@ifundefined{chapter}{\def\thebibliography#1{
\section*{References}
  \list
  {\relax}{\setlength{\labelsep}{0em}
        \setlength{\itemindent}{-\bibhang}
        \setlength{\itemsep}{\parskip}
        \setlength{\parsep}{0pt}
        \setlength{\leftmargin}{\bibhang}}
    \def\newblock{\hskip .11em plus .33em minus .07em}
    \sloppy\clubpenalty4000\widowpenalty4000
    \sfcode`\.=1000\relax}}%
{\def\thebibliography#1{
  \list
  {\relax}{\setlength{\labelsep}{0em}
        \setlength{\itemindent}{-\bibhang}
        \setlength{\itemsep}{\parskip}
        \setlength{\parsep}{0pt}
        \setlength{\leftmargin}{\bibhang}}
    \def\newblock{\hskip .11em plus .33em minus .07em}
    \sloppy\clubpenalty4000\widowpenalty4000
    \sfcode`\.=1000\relax}}
 
\newlength{\bibhang}
\let\@internalcite\cite
\def\cite{\@ifstar{\citey}{\citefull}}
\def\citefull{\def\astroncite##1##2{##1\ ##2}\@internalcite}
\def\citey{\def\astroncite##1##2{##1\ (##2)}\@internalcite}
\def\citeyear{\def\astroncite##1##2{##2}\@internalcite}
\def\citename{\def\astroncite##1##2{##1}\@internalcite}
\def\@citex[#1]#2{\if@filesw\immediate\write\@auxout{\string\citation{#2}}\fi
  \def\@citea{}\@cite{\@for\@citeb:=#2\do
    {\@citea\def\@citea{; }\@ifundefined
       {b@\@citeb}{{\bf ??}\@warning
       {Citation `\@citeb' on page \thepage \space undefined}}%
{\csname b@\@citeb\endcsname}}}{#1}}
 
\def\@cite#1#2{#1\if@tempswa #2\fi} 
\def\@biblabel#1{}
 
\def\astroncite#1#2{#1\ #2}
\makeatother

\begin{document}

\title{An {\it RXTE} study of M87 and the core of the Virgo cluster}

\author{Christopher S. Reynolds\altaffilmark{1,2}, Sebastian
  Heinz\altaffilmark{2} and Andrew C. Fabian\altaffilmark{3}}

 
\altaffiltext{1}{Hubble Fellow}

\altaffiltext{2}{JILA, University of Colorado, Campus Box 440, Boulder, CO 80309-0440\\
\{chris,heinzs\}@rocinante.colorado.edu}

\altaffiltext{3}{Institute of Astronomy, Madingley Road, Cambridge CB3~OHA,
  UK}
 

\begin{abstract}
  We present hard X-ray observations of the nearby radio galaxy M87 and the
  core of the Virgo cluster using the {\it Rossi X-ray Timing Explorer}.
  These are the first hard X-ray observations of M87 not affected by
  contamination from the nearby Seyfert 2 galaxy NGC~4388.  Thermal
  emission from Virgo's intracluster medium is clearly detected and has a
  spectrum indicative of $kT\approx 2.5\keV$ plasma with approximately 25\%
  cosmic abundances.  No non-thermal (power-law) emission from M87 is
  detected in the hard X-ray band, with fluctuations in the Cosmic X-ray
  Background being the limiting factor.  Combining with {\it ROSAT} data,
  we infer that the X-ray spectrum of the M87 core and jet must be steep
  ($\Gamma_{\rm core}>1.90$ and $\Gamma_{\rm jet}>1.75$), and we discuss
  the implications of this result.  In particular, these results are
  consistent with M87 being a mis-aligned BL-Lac object.
\end{abstract}

\section{Introduction}

The nearest giant elliptical galaxy, M87 (NGC~4486), holds a central place
in the study of low-luminosity radio galaxies and extragalactic radio jets.
This galaxy, situated at the center of the Virgo cluster of galaxies, is
associated with the Faranoff-Riley class I radio source Virgo-A and
displays the most prominent extragalactic radio jet in the northern sky.
It was the first extragalactic jet to be discovered (Curtis 1918) and has
since been subjected to intense observational study at all available
wavelengths (see Biretta 1993 for a review).  The close proximity of this
source, about 16\,Mpc (e.g., Tonry 1991), makes it a crucial laboratory for
testing our understanding of both extragalactic jets and the central engine
structure of radio-loud AGN.

In the soft X-ray band, imaging with the high-resolution imagers (HRIs) on
both the {\it Einstein} and {\it ROSAT} satellites have resolved emission
from the core of M87 and knots A, B and D of its optical jet (Schreier,
Gorenstein \& Feigelson 1982; Biretta, Stern \& Harris 1991; hereafter
BSH91).  The mechanisms underlying any of these emission components is
unknown.  Suggestions for the jet emission mechanism include synchrotron
emission from ultra-relativistic ($\gamma\sim 10^7$) electrons in the jet
plasma, inverse Compton scattering of infra-red/optical photons by a
population of $\gamma\sim 100$ electrons in the jet plasma, and thermal
bremsstrahlung from shock heated gas surrounding the jet.  The observed
core emissions could represent the inner jet with one of the above
mechanisms producing the X-rays.  On the other hand, emission from an
accretion disk corona (as in the Seyfert case) or a hot accretion disk
(such as an Advection Dominated Accretion Flow; Reynolds et al. 1996) might
also be important for understanding the core emissions.

For years there was a mystery surrounding the hard X-ray emission from M87
and the Virgo cluster.  Whereas the centroid of the low-energy emission
lies on M87, Davison (1978) used {\it Ariel-V} data to show that the
higher-energy emissions possessed a different centroid (displaced to the
north-west by $\sim 1\degmark$).  This puzzle was resolved by the
coded-mask imaging from {\it Spacelab-2} which found that the high-energy
emissions ($10\keV$ or greater) of the Virgo cluster are dominated by the
Seyfert 2 galaxy NGC~4388 (Hanson et al. 1990).  NGC~4388 is displaced from
M87 by just over a degree to the north-west.  An unambiguous measurement of
the hard X-ray flux, spectrum and variability properties of M87 has proven
difficult due to the presence of this confusing source.  Takano \& Koyama
(1991) analyzed {\it Ginga} scanning data and determined a photon index of
$\Gamma=1.9\pm 0.02$ and a 10--20\,keV flux of $F_{10-20\keV}=1.6\times
10^{-11}\ergpcmsqps$, corresponding to a 2--10\,keV flux of
$F_{2-10\keV}=3.4\times 10^{-11}\ergpcmsqps$ (assuming a simple
extrapolation of the 10--20\,keV powerlaw to lower energies).  Hanson et al.
(1990) report upper limits that are 5 times weaker.  Takano \& Koyama
(1991) take this as evidence for variability.

With the exception of the {\it EXOSAT} ME which suffered from severe
background subtraction issues at high energies, the {\it Rossi X-ray Timing
  Explorer (RXTE)} is the first hard X-ray observatory with a sufficiently
small field of view to avoid contamination by NGC~4388.  In this {\it
  letter} we report four {\it RXTE} observations of M87.  The AGN/jet is
not detected above the thermal emission of the Virgo cluster, and upper
limits on its flux are given.  We include the effect of unknown
fluctuations in the Cosmic X-ray Background (CXB) when deriving our limits
on the non-thermal emission and these, indeed, turn out to be the limiting
factor for these data.  Astrophysical implications of this non-detection
are discussed in Section 4.

\section{Observations and basic data reduction}

\begin{figure*}
\centerline{\psfig{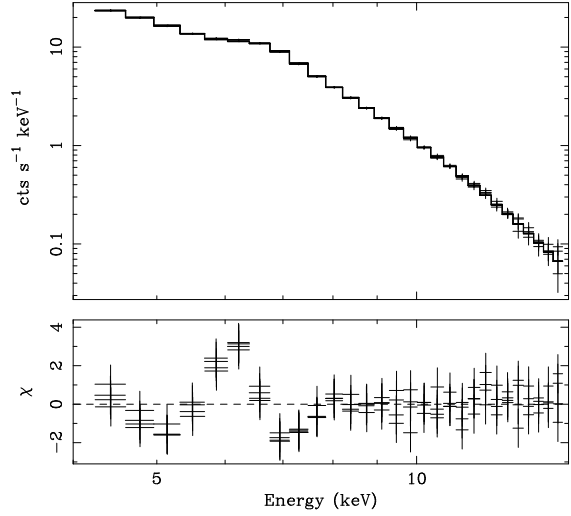}}
\caption{The best fitting two-component thermal fit to the combined
  PCA data.}
\end{figure*}

We observed M87 with {\it RXTE} four times.  The dates (and good on-source
exposure times) were 30-Dec-1997 to 2-Jan-1998 (42600\,s), 9-Jan-1998 to
12-Jan-1998 (43500\,s), 19-Jan-1998 to 22-Jan-1998 (35500\,s), and
30-Jan-1998 to 3-Feb-1998 (44900\,s).  The motivation of the project was to
search for the hard X-rays from M87 and its jet, and study their spectrum
and variability.

{\it RXTE} is a hard X-ray observatory possessing two pointed instruments,
the Proportional Counter Array (PCA) and the High Energy X-ray Timing
Experiment (HEXTE), as well as an all-sky monitor (ASM).  In this work, we
use data from the PCA and HEXTE.

The PCA consists of five nearly identical co-aligned Xenon proportional
counter units (PCUs) with a total effective area of about 6500\,cm$^{2}$
and is sensitive in the energy range from 2\,keV to $\sim 60$\,keV (Jahoda
et al.  1996).  Data taken in {\sc Standard-2} mode were extracted into
spectra and lightcurves using {\sc Ftools v4.1} supplemented with the {\it
  RXTE} patch (A. Smale, private communication).  For spectral fitting,
response matrices were generated using the {\sc ftools} routine {\sc pcarmf
  v3.3} (corrected for the 1998-Aug-29 bug in which {\sc pcarmf} failed to
account for temporal variations of the response matrix).  To take into
account the remaining uncertainties of the matrix we added 1\% systematic
errors to our data.  This value was determined from the deviations from a
pure power-law in a fit to the Crab Nebula and pulsar spectrum.  Background
subtraction was performed using the latest background models (released in
June 1998 as part of the RXTE patch to {\sc ftools v4.1}).  To increase
signal-to-noise, only data from the top layer of the PCUs are considered
here.  We limit our consideration to the energy range 4--15\,keV --- the
lower bound is determined by the lower limit of the well calibrated
energies, whereas the upper bound is the energy at which the data become
background dominated.

The HEXTE consists of two clusters of four NaI/CsI-phoswich scintillation
counters, sensitive from 15 to 250\,keV (Rothschild et al. 1998).
Background subtraction is done by source-background rocking of the two
clusters.  No signal from M87 was detected in HEXTE for either the
individual observations or the co-added data of all of our datasets.
However, the limits set by these non-detections are uninteresting compared
with the PCA limits that we shall address below, and hence we shall not
discuss HEXTE data any further.

The total background subtracted PCA count rate was $113\,{\rm counts}\,{\rm
  s}^{-1}$ for all 5 PCUs, with no evidence for variability either
between the four observations or within individual observations.  In the
absence of any temporal structure, we focus on the spectral aspects.  We
have extracted 4 PCA spectra, one for each individual observation, and
rebinned to at least 20 photons per energy bin.  This is a requirement for
the $\chi^2$ analysis of the following section to be appropriate.  Spectral
fitting was then performed using the {\sc xspec v10.0} package.

\section{Spectral results}

In this section, the 4--15,keV spectra resulting from the four observations
are analyzed individually in order to assess spectral variability.  Since
we fail to detect spectral variability, we also fit the combined spectrum
jointly.  We expect each spectrum to be dominated by thermal plasma
emission from the ICM of the Virgo cluster.  Non-thermal emission from the
AGN or jet, if present, would be revealed as a hard tail above and beyond
the thermal emission.

For these data, unknown fluctuations in the CXB are the limiting factor in
our ability to study the non-thermal emission from M87.  The spectral
fitting presented in section 3.1 (and in Table~1) incorporates this
uncertainty by including such fluctuations as an extra model component.  In
detail, we include a power-law component with photon index $\Gamma=1.8$
which is allowed to vary in normalization (measured at 1\,keV) between
$N_{\rm CXB}=\pm 2\times 10^{-4}\,{\rm ph}\,{\rm keV}^{-1}\,{\rm
  cm}^{-2}\,{\rm s}^{-1}$ (i.e., the 1-$\sigma$ fluctuations of the CXB
estimated by scaling the {\it Ginga} results of Butcher et al. [1997] to
account for the instrumental responses, effective areas and fields of
view).

\subsection{The thermal plasma and limits on a $\Gamma=2$ power-law}

Initially, these spectra were fitted with a single temperature thermal
plasma model (Mewe, Gronenschild, \& van den Oord 1985; Kaastra 1992)
modified by Galactic absorption ($N_{\rm H}=2.5\times 10^{20}\pcmsq$) and
redshifted appropriately for the Virgo/M87 system ($z=0.003$).  The
resulting fits are acceptable ($\chi^2_\nu\sim 1.1$ for 25 degrees of
freedom) and are shown in Table~1.  As can be seen from Table~1, all
spectra are adequately described by a $kT\approx 2.5\keV$ plasma with an
abundance of $Z\approx 0.26\,Z_\odot$.  Extrapolating the fits in the
2--4\,keV range, the total inferred 2--10\,keV flux is $2.8\times
10^{-10}\ergpcmsqps$, although there may well be a cooler cluster component
contributing to the cluster X-ray emission below 4\,keV (e.g. Lea et al.
1982; Matsumoto et al. 1996).  The corresponding 2--10\,keV luminosity is
$8.7\times 10^{42}\ergps$.

The results of the thermal plasma fit to all four datasets simultaneously
(i.e., the combined fit) are shown in Fig~1.  A `line-like' feature can be
seen at $\sim 6\keV$.  This feature is seen in all PCU detectors
individually and in each observing period.  An additional narrow Gaussian
line with energy $E=6.22\pm 0.12\keV$, slightly less than the cold iron
fluorescent line energy of iron, and equivalent width $84\eV$ describes
this feature well and produces a large improvement in the goodness of fit
($\chi^2_\nu=56.0/107$).  There are several reasons for suspecting that
this line is {\it not} real.  Firstly, one must be immediately suspicious
since this is a weak line-feature embedded in the wings of a much stronger
line (i.e., the ionized emission line from the thermal cluster gas).
Secondly, such a line is not seen in {\it ASCA} data of the Virgo cluster
(an examination of archival {\it ASCA} data gives an upper limit of $\sim
20\eV$ on the equivalent width of a line at these energies).  Thirdly,
there is no astrophysical precedent for observing such a line from a
cluster dominated system.  While we cannot firmly reject the hypothesis
that this line is real, we suspect that it is due to a small mismodelling
of the ionized emission lines.  Furthermore, the presence of this line in
the spectrum does not affect the best fit values or uncertainties of the
other spectral parameters. Thus, we shall not discuss this feature any
further, and shall not include it in the spectral fitting described below.

\begin{table*}
\begin{center}
\begin{tabular}{cccccc}\hline
Model \& & Interval & Interval & Interval & Interval \\
model parameters & 1 & 2 & 3 & 4 & combined \\\hline
NH+MEKA+BACK & \\
$kT$ (keV) & $2.57^{+0.14}_{-0.15}$ & $2.56\pm 0.03$ &
$2.54^{+0.03}_{-0.04}$ & $2.54^{+0.02}_{-0.03}$ & $2.54^{+0.02}_{-0.01}$ \\
$Z$ ($Z_\odot$) & $0.26^{+0.01}_{-0.02}$ & $0.26^{+0.02}_{-0.01}$ & $0.26\pm 0.02$ &
$0.26^{+0.02}_{-0.01}$ & $0.26\pm 0.01$ \\
$N$ (${\rm photons}\,{\rm s}^{-1}\,{\rm cm}^{-2}\,{\rm keV}^{-1}$ @ 1keV) & $0.62^{+0.01}_{-0.02}$
& $0.62^{+0.01}_{-0.02}$ & $0.63^{+0.02}_{-0.01}$ & $0.63\pm 0.02$ &
$0.63\pm 0.01$ \\
$\chi^2$/dof & 28.6/25 & 27.4/25 & 27.4/25 & 27.2/25 & 120/109 \\\hline
NH+MEKA+PO+BACK & \\
$N_{\rm pow}$ ($10^{-3}\,{\rm photons}\,{\rm
    s}^{-1}\,{\rm cm}^{-2}\,{\rm keV}^{-1}$ @ 1keV)& $<1.8$ & $0.9^{-0.4}_{+1.0}$ & $<1.2$ & $<1.9$ & $<1.2$ \\ 
$F_{\rm pow}(2-10\keV)$ ($10^{-12}\ergpcmsqps$) & $<4.6$ &
$<2.3^{+2.6}_{-1.0}$ & $<3.1$ & $<4.6$ & $<3.1$ \\
$\chi^2$/dof & 27.2/24 & 24.1/24 & 27.3/24 & 26.6/24 & 118/108 \\\hline
\end{tabular}
\caption{Spectral fits to {\it RXTE}-PCA data in 4--15\,keV range.  $N_{\rm
    pow}$ and $N$ are the normalizations of the power-law component and the
  thermal component, respectively, at 1\,keV.  The power-law is assumed to
  have a photon index of $\Gamma=2$ in these fits.  All errors are quoted
  at the 90 per cent confidence level for one interesting parameter
  ($\Delta\chi^2=2.7$).}
\end{center}
\end{table*}

In order to assess the presence of hard non-thermal emission from the AGN
or jet, a power-law component was added to the thermal plasma model.  In no
case did the addition of the power-law component lead to any significant
improvement in the goodness of fit.  Table~1 quotes the 90 per cent limits
on the normalization of the power law (at $1\keV$) assuming a photon index
of $\Gamma=2$ (close to the canonical value for Type-1 AGN).

\subsection{Detailed limits on non-thermal emission}

Given the lack of any spectral variability, we choose to refine our limits
on the non-thermal emission using the combined spectrum (i.e., summing the
spectra from our individual observations).  Here, we compute confidence
contours on the $(\Gamma,N_{\rm pow})$-plane using a more rigorous
treatment of the effect of CXB fluctuations on our spectral fitting.

Since we know the probability distribution of the CXB fluctuations (Butcher
et al. 1997), we can integrate over possible fluctuations and obtain
`averaged' confidence contours.  We define a likelihood function given a
particular CXB fluctuation,
\begin{equation} 
{\cal L}(\Gamma,N_{\rm pow}|N_{\rm CXB})\propto \exp\left(-\frac{\chi^2(\Gamma,N_{\rm pow}|N_{\rm CXB})}{2}\right),
\end{equation}
where $\chi^2(\Gamma,N_{\rm pow}|N_{\rm CXB})$ is derived from fitting the
spectral model of the previous section to the combined data with $\Gamma,
N_{\rm pow}$ and $N_{\rm CXB}$ fixed at given values.  Defining the
probability of a given CXB fluctuation as $p(N_{\rm CXB})$, we can
integrate the likelihood function over this parameter,
\begin{equation} 
{\cal L}(\Gamma,N_{\rm pow})=\int_{-\infty}^{+\infty}{\cal L}(\Gamma,N_{\rm
  pow}|N_{\rm CXB})\,p(N_{\rm CXB})\,{\rm d}N_{\rm CXB}.
\end{equation}
Operationally, we compute the $\chi^2$ surfaces over the $(\Gamma,N_{\rm
  pow})$-plane with given values of $N_{\rm CXB}$ using the {\sc xspec}
package (as would be done if we were to compute regular confidence
contours).  Given these $\chi^2$ surfaces, we approximate eqn (2) by a
simple sum weighted according to the Gaussian probability function
$p(N_{\rm CXB})$.  

From ${\cal L}$, we can produce confidence contours on the $(\Gamma,N_{\rm
  pow})$-plane in which the CXB fluctuations have been averaged over.
Figure~2 shows the resulting 68\%, 90\%, and 95\% contours.

\begin{figure*}[t]
\centerline{\psfig{figure=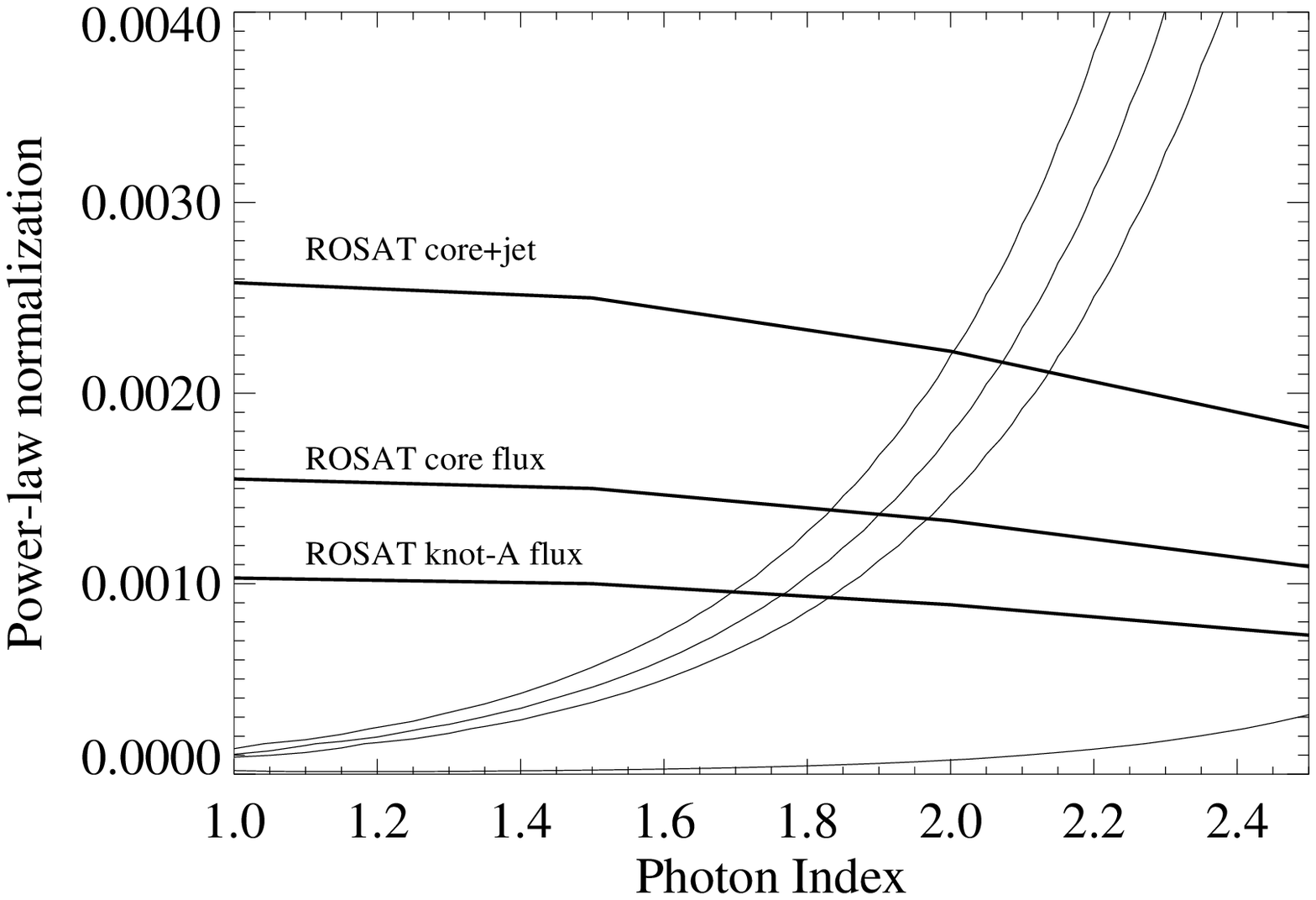,width=0.5\textwidth}}
\caption{Confidence contours on the $\Gamma$-$N_{\rm
    pow}$ plane.  Parameter space to the top left of the plane is
  inconsistent with these data.  From the top to the bottom, the contours
  mark the 95\%, 90\% and 68\% (two contours) confidence levels for two
  interesting parameters. Also shown (as thick lines) are the lines in
  parameter space that are consistent with the Jan-1998 {\it ROSAT} data of
  Harris, Biretta \& Junor (1998), assuming that a single power-law form
  exists from the {\it ROSAT}-band to the {\it RXTE} band (modified by
  Galactic absorption of $N_{\rm H}=2.5\times 10^{20}\pcmsq$).}
\end{figure*}

The resulting upper limit on the hard X-ray power-law from M87 assuming
$\Gamma=2$ is
\begin{equation}
F_{\rm pow}(2-10\keV)<4.1\times 10^{-12}\ergpcmsqps,
\end{equation}
corresponding to an isotropic luminosity of
\begin{equation}
L_{\rm pow}(2-10\keV)<1.0\times 10^{41}\ergps.
\end{equation}
using a distance of 16\,Mpc.  These limits are quoted at the 90 per cent
confidence level.

\section{Discussion}

Our upper limit on the hard X-ray power-law component of M87 are the most
stringent to date.  They are superior to the upper limits derived from {\it
  Spacelab-2} data [$F_{\rm pow}(2-10\keV)<8\times 10^{-12}\ergpcmsqps$;
Hanson et al. 1990] and {\it ASCA} data [$F_{\rm pow}(2-10\keV)<8\times
10^{-12}\ergpcmsqps$; Reynolds et al. 1996].  Note that Matsumoto et al.
(1996) claim a detection of the M87 power-law with $F_{\rm
  pow}(2-10\keV)\approx 8\times 10^{-12}\ergpcmsqps$ using the same {\it
  ASCA} data as Reynolds et al.  (1996).  However, the complex structure of
the thermal ICM, which possesses temperature and abundance gradients as
well as a possible multiphase structure, makes the thermal ICM spectrum
difficult to model and so renders such conclusions about superposed
non-thermal emission open to suspicion.  Our limits are also inconsistent
with the value of $F_{\rm pow}(2-10\keV)\approx 3.4\times
10^{-11}\ergpcmsqps$ derived from {\it Ginga} scanning data (Tanako \&
Koyama 1991).  While this may be evidence for variability, it is also
possible that scattered flux from NGC~4388 coupled with the complex thermal
ICM influences the subtle analysis of Tanako \& Koyama (1991).  In summary,
we would argue that there has never been an irrefutable detection of a hard
X-ray power-law from M87.

However, both the {\it Einstein} and {\it ROSAT} HRIs have imaged variable
soft X-ray point sources coincident with the core of M87 and knot-A
(Schreier, Gorenstein \& Feigelson 1982; BSH91; Harris, Biretta \& Junor
1998; hereafter HBJ98).  The last {\it ROSAT} measurement shown by HBJ98
was taken on 5-Jan-1998 and hence lies within these {\it RXTE}
observations.  Assuming a single (absorbed) power-law form extending from
the {\it ROSAT} band into the {\it RXTE} band, the 1998 {\it ROSAT} fluxes
can be converted into loci on the $\Gamma$-$N_{\rm pow}$ plane, as shown in
Fig.~2.  It can be seen that the {\it ROSAT} sources must possess a fairly
steep high-energy spectrum ($\Gamma\approxgt 1.90$ and $\Gamma\approxgt
1.75$ for the core and jet respectively) in order to avoid detection by
{\it RXTE}.  If the core and jet components have the same hard X-ray
spectrum, then the spectral slope must exceed $\Gamma>2.05$ in order for the
combined hard X-rays from these two sources to remain undetected.  If there
is significant intrinsic absorption in this system, even steeper intrinsic
spectra are required.

It is informative to compare these limits with the photon indices found in
various classes of AGN.  In a recent {\it ASCA} study, Reeves et al. (1997)
found that radio-loud quasars possess {\it ASCA} band photon indices of
$\Gamma=1.63\pm 0.04$ (see also studies by Lawson et al. 1992, and Williams
et al. 1992).  Our data rule out the possibility that the {\it ROSAT}
source seen in the core of M87 possesses such a flat X-ray spectrum when
extrapolated to the {\it RXTE} band.  However, Tsvetanov et al. (1998) have
recently suggested that M87 is a misaligned example of a BL-Lac object.
These AGN typically possess steep {\it ASCA}-band X-ray spectra with
$\Gamma\sim 1.8$ for the low-energy peaked BL-Lacs (LBL) and $\Gamma\sim
2-3$ for the high-energy peaked BL-Lacs (HBL; e.g. Kubo et al. 1998).  The
lack of an observed break in the optical-UV spectrum of the jet (Tsvetanov
et al. 1998), and the steep X-ray spectral index, would suggest that it
belong to the HBL catagory.  Thus, our data are entirely consistent with
the suggestion that M87 is a misaligned HBL.

\section{Conclusions}

We have presented four {\it RXTE} observations of M87 spread over the month
of Jan-1998 and totaling approximately 167\,ksec of on-source exposure
time.  Thermal plasma emission from the Virgo cluster ICM is clearly
detected in the PCA with a temperature of $kT\approx 2.5\,keV$.  The metal
abundance of the plasma is inferred to be $Z\approx 0.26\,{\rm Z}_\odot$,
although this must be treated as an average abundance given the known
abundance gradients in this system (Matsumoto et al. 1996).  There was no
detection in the HEXTE.

Once the thermal ICM emission has been modeled, there is no detection of
any hard X-ray non-thermal (power-law) emission in the PCA spectrum.  Our
upper limit on the flux of a $\Gamma=2$ power-law component is $F_{\rm
  pow}<4.1\times 10^{-12}\ergpcmsqps$.  Fluctuations in the CXB are the
limiting factor in our ability to set upper limits on the non-thermal
emission.  If the core and jet sources detected by {\it ROSAT} possess a
power-law spectrum into the {\it RXTE} band, the photon indices of this
sources must be $\Gamma_{\rm core}>1.90$ and $\Gamma_{\rm jet}>1.75$
respectively.  This is entirely consistent with the hypothesis that M87 is
a misaligned HBL.

\section*{Acknowledgments}

We thank Julia Lee, Michael Nowak, Beverly Smith, and J\"orn Wilms for
useful discussions.  We also thank Andrew Hamilton for sharing his
statistical expertise, and Craig Sarazin for pointing out an error in a
previous draft of this manuscript.  CSR thanks support from NASA under LTSA
grant NAG5-6337.  CSR also thanks support from Hubble Fellowship grant
HF-01113.01-98A awarded bythe Space Telescope Institute, which is operated
by the Association of Universities for Research in Astronomy, Inc., for
NASA under contract NAS 5-26555.  ACF thanks the Royal Society for support.


\small
\begin{thebibliography}{}
\bibitem[]{} Biretta J.~A., 1993, in {\it Astrophysical Jets}, p. 263, eds.
  Burgarella D., Livio M., O`Dea C.~P.,Cambridge University Press,
  Cambridge 
\bibitem[]{} Biretta J.~A., Stern C.~P., Harris D.~E., 1991, AJ, 101, 1632 (BSH91)
\bibitem[]{} Butcher J.~A. et al., 1997, MNRAS, 291, 437
\bibitem[]{} Curtis H.~D., 1918, Pub. Lick Obs., 13, 31
\bibitem[]{} Davison P.~J.~N., 1978, MNRAS, 183, 39
\bibitem[]{} Ford H.~C. et al. 1995, ApJ, 1994, 435, L27
\bibitem[]{} Jahoda K., Swank J.~H., Giles A.~B., Stark M.~J., Strohmayer
  T., Zhang W., Morgan E.~H., 1996, in EUV, X-ray, and Gamma-Ray
  Instrumentation for Astronomy VII, ed. O.~H.~Siegmund, (Bellingham, WA:
  SPIE), 59
\bibitem[]{} Hanson C.~G., Skinner G.~K., Eyles C.~J., Willmore A.~P.,
  1990, MNRAS, 242, 262
\bibitem[]{} Harris D.~E., Biretta J.~A., Junor W., 1998, The M87 Ringberg
  Workshop, eds H.~J.~Roser \& K.~Meisenheimer, Spinger (astrop-ph/9804201;
  HBJ98) 
\bibitem[]{} Kaastra J.~S.~, 1992, An X-Ray Spectral Code for Optically Thin
  Plasmas (Internal SRON-Leiden Report, updated version 2.0)
\bibitem[]{} Kubo H., Takahashi T., Madejski G., Tashiro M., Makino F.,
  Inoue S., Takahara F., 1998, ApJ, 504, 693 
\bibitem[]{} Lawson A.~J., Turner M.~J.~L., Williams O.~R., Stewart G.~C.,
  Saxton R.~D., 1992, MNRAS, 259, 743 
\bibitem[]{} Lea S.~M., Mushotzky R.~F., Holt S.~S., 1982, ApJ, 262, 24
\bibitem[]{} Matsumoto H., Koyama K., Awaki H., Tomida H., Tsuru T.,
  Mushotzky R.~F., Hatsukade I., 1996, PASJ, 48, 201 
\bibitem[]{} Mewe R., Gronenschild E.~H.~B.~M., van den Oord G.~H.~J., 1985,
       A\&AS, 62, 197
\bibitem[]{} Reeves J.~N., Turner M.~J.~L., Ohasi T., Kii T., 1997, MNRAS,
  292, 468
\bibitem[]{} Reynolds C.~S., Di~Matteo T., Fabian A.~C., Hwang U., Canizares
C.~R., 1996, MNRAS, 283, L111
\bibitem[]{} Rothschild R.~E. et al., 1998, ApJ, 496, 538
\bibitem[]{} Schreier E.~J., Gorenstein P., Feigelson E.~D., 1982, ApJ,
  261, 42
\bibitem[]{} Takano S., Koyama K., 1991, PASJ, 43, 1
\bibitem[]{} Tonry J.~L., 1991, ApJ, 373, L1
\bibitem[]{} Tsvetanov Z.~I. et al., 1998, ApJ, 493, L83
\bibitem[]{} Williams O.~R. et al., 1992, ApJ, 389, 157
\end{thebibliography}
\end{document}